\documentstyle[12pt]{article}
\newcommand{\dslash}{\not\!\partial}
\newcommand{\Aslash}{\not\!\! A}
\newcommand{\sign}{\rm sign}

\advance\textheight by \topskip

\begin{document}
\mbox{}
\begin{flushright}
UM-P-94/58\\
RCHEP-94/16\\
May, 1994
\end{flushright}
%
%
\begin{center}
{\Large\bf What's wrong with anomalous chiral gauge
theory?\footnote{Talk given at the First International Symposium
on Symmetries in Subatomic Physics, Taipei, 16-18 May, 1994.}}

\vspace{.1in}

Tien D Kieu,\\
School of Physics and Research Centre for High Energy Physics,\\
University of Melbourne,\\
Parkville VIC 3052,\\
AUSTRALIA

\vspace{.3in}
{\bf Abstract}
\end{center}
\begin{quotation}
\noindent
It is argued on general ground and demonstrated in the particular
example of the Chiral Schwinger Model that there is nothing wrong
with apparently anomalous chiral gauge theory.  If quantised correctly,
there should be no gauge anomaly and chiral gauge theory should be
renormalisable and unitary, even in higher dimensions and with
non-abelian gauge groups.
Furthermore, mass terms for gauge bosons and chiral fermions can be
generated without spoiling the gauge invariance.
\end{quotation}
\section*{Introduction}
Symmetry has long been an important concept in physics and even broken
symmetry is neither less important nor less useful.  In particle physics,
symmetry is the underlying principle of gauge field theory wherein
all fundamental interactions are described.

In addition to explicit or spontaneous breaking,
quantum fluctuations can also destroy a symmetry which is exact at the
classical level.  Such breaking was a surprise and hence named anomaly.
The well-known anomaly
of global axial symmetry is due to the need to regulate short distance
singularity of the kind of quantum fluctuations associated with all Feynman
loop diagrams~\cite{abj}.
This kind of anomaly has found applications in the
phenomenology of neutral pion decays, in the $U(1)$ problem and possibly
in the proton spin structure.

On the other hand, the anomalous breaking of the local symmetry of a
gauge theory is undesirable.  It is logically inconsistent if the
local gauge invariance which is the underlying principle and also the
definition of a gauge theory is lost upon quantisation.  Indeed, the
inconsistency does show up in the mathematical formulation of anomalous
gauge theory.

Take the example of an abelian chiral gauge theory where only the
left-handed fermion current is coupled to the gauge fields,
\begin{eqnarray}
{\cal L} &=& -\frac{1}{4}F^2 + \bar\psi(i\dslash + g\Aslash P_L)\psi.
\label{1}
\end{eqnarray}
$P_{L,R} = \frac{1}{2}(1\mp\gamma_5)$ are the chirality projectors.
This Lagrangian density is invariant under the gauge transformations
\begin{eqnarray}
gA_\mu(x) &\to& gA_\mu(x) + \partial_\mu\theta(x),\nonumber\\
\psi(x) &\to& [\exp(i\theta(x))P_L + P_R]\psi(x),\\
\bar\psi(x) &\to& \bar\psi(x)[P_L + \exp(-i\theta(x))P_R]\nonumber.
\label{2}
\end{eqnarray}

Upon quantisation along the conventional route, it can be shown,
via either Feynman diagrams~\cite{abj} or the path-integral
jacobean~\cite{fujikawa}, that the gauged current
\begin{eqnarray}
J^\mu_L &=& \bar\psi\gamma^\mu P_L\psi
\label{3}
\end{eqnarray}
is not conserved, as in two dimensions (with $\epsilon^{01}=1$)
\begin{eqnarray}
\partial_\mu J^\mu_L &=& \frac{\hbar g}{4\pi}\epsilon^{\mu\nu}
\partial_\mu A _\nu.
\label{4}
\end{eqnarray}
Explicit $\hbar$ is inserted to remind us that this is a quantum effect.

The gauge symmetry is then lost, destroying the Ward-Takahashi identities
which are instrumental in the consideration of perturbative
renormalisability and unitarity.

Furthermore, the anomaly~(\ref{4}) is not compatible with the equation
of motion, $\partial_\mu F^{\mu\nu} = gJ_L^\nu$,
\begin{eqnarray}
0 = \partial_\mu\partial_\nu F^{\mu\nu} &=& g\partial_\mu
 J^\mu_L \not= 0.
\label{5}
\end{eqnarray}
The Gauss-law constraint also changes its nature from first-class to
second-class when the commutator of Gauss-law operators is no longer
non-vanishing.
This transmutation takes place {\it after} quantisation and consequently
entails contradictions both in the quantisation procedure, which requires
different treatments for different types of constraints, and in the
imposition of the constraints on the Hilbert space.  To make it even harder
to be enforced, the constraint is now time dependent~\cite{4} since it is
no longer the generator of a symmetry and thus does not commute with the
Hamiltonian.

Last but neither least nor separately, chiral gauge theory, anomalous
or non-anomalous, does not seem to admit any
non-perturbative definition.  In Lattice Gauge Theory, in particular,
there are un-surmountable difficulties to implement chiral fermions.

All the above, except the last difficulty of the last paragraph,
leads to the requirement that gauge anomaly is to be cancelled.
The accepted wisdom has the cancellation in appropriate choice of
gauge groups or fermion representation, as with the conspiracy of
quarks and leptons in the standard electroweak model.

Notwithstanding this, I take the approach that gauge anomaly and its
problems are perhaps due to our ignorance.  I have advocated
elsewhere~\cite{tdk-berry,tdk-prd,tdk-mp} and will substantiate in
an example below that
the conventional quantisation of chiral gauge theory is wrong.  The
proposition is that if the quantisation is done correctly, a chiral gauge
theory should:
\begin{itemize}
\item have no gauge anomaly, but anomaly of {\it global} symmetry is still
admissible;
\item be perturbative renormalisable;
\item be unitary;
\item lead to a solution of the problem of lattice chiral fermions;
\item and as a bonus, be able to admit gauge-invariant mass for both
gauge bosons and chiral fermions without the need for scalar field.
\end{itemize}

In the next section, I review a classic example of how a theory might
be quantised incorrectly.  Following this is a section of the pitfalls to
be aware of in the second quantisation.  The example
of two-dimensional abelian chiral gauge theory, known as the Chiral Schwinger
Model, in the following section illustrates some of the consistency claims
above for chiral gauge theory.  The last section of the paper is for some
concluding remarks.
\section*{Correct quantisation}
Correct quantisation is synonymous with correct Feynman rules in perturbation
context or correct path-integral action in general.  Richard Feynman himself
was probably the first to run into inconsistency with incomplete set of
Feynman rules in a non-abelian gauge theory~\cite{feynman}.  If loop
diagrams involving two or more gluons in a given gauge are constructed
according to the simple rules for tree diagrams, that is, from classical
action as opposed to path-integral action, unitarity will be violated.
Feynman then recognised that ghosts, scalar
particles with wrong statistics but not showing up in asymptotic states,
were necessary to restore unitarity.  However, this {\it ad hoc} treatment
only works at one-loop level because unitarity is not a sufficient
constraint to treat more than one gluon loop.

These ghosts are now known as Faddeev-Popov ghosts and are shown in
the path-integral framework to be necessary to restore unitarity to all
orders in perturbation theory.  The introduction of ghosts
brings some degrees of non-locality to the path integral.  But even so,
the theory is still considered to be local in the sense that space-like
commutators of physical fields vanish.  I will come back to this
{\it apparent} non-locality later on in the treatment of chiral gauge
theory.

The lesson we have learned from this example is that apparent
inconsistency might be due to the use of erroneous or incomplete
Feynman rules.  And correct Feynman rules are not those naively derived
from the classical action.  This is the central theme of the paper
as I will argue that similar ignorance could lead to
problems in a general chiral gauge theory.
\section*{Review of second quantisation}
In the interaction picture, one introduces an evolution operator,
$U(t,t_0)$, to carry a state from time $t_0$ to time $t$.  On the
other hand, all the field operators are free-field operators, that is,
they are evolving in time according to the non-interaction part of the
Hamiltonian.  And the $S$ matrix can be obtained in the limit
\begin{eqnarray}
S &=& \lim_{t\to\infty}\lim_{t_0\to -\infty} U(t,t_0).
\label{6}
\end{eqnarray}

The Hilbert-space state evolution operators satisfies the
following Tomonaga-Schwinger equation
\begin{eqnarray}
\partial_tU(t,t_0) &=& -iH_{int}(t)U(t,t_0),
\nonumber\\
U(t_0,t_0) &=& 1,
\label{7}
\end{eqnarray}
where $H_{int}(t)$ is the interaction part of the Hamiltonian.
Usually, the solution for the evolution operator is given in the
form,
\begin{eqnarray}
\tilde U(t,t_0) &=& {\cal T}\exp\left\{-i\int^t_{t0}
H_{int}(\tau)d\tau\right\}, \nonumber\\
&=& 1 + (-i)\int^t_{t_0}H_{int}(\tau)d\tau + \frac{(-i)^2}{2!}\int^t_{t_0}
\int^t_{t_0}{\cal T}[H_{int}(\tau)H_{int}(\tau')]d\tau d\tau'
+ \cdots
\label{8}
\end{eqnarray}
$\cal T$ is the Dyson time ordered product,
\begin{eqnarray}
{\cal T}[A(\tau)B(\tau')] &=& \theta(\tau-\tau')A(\tau)B(\tau')
+ \theta(\tau'-\tau)B(\tau')A(\tau),\nonumber\\
&=& \frac{1}{2}\left\{A(\tau),B(\tau')\right\}
+ \frac{1}{2}\sign(\tau-\tau')\left[A(\tau),B(\tau')\right].
\label{8a}
\end{eqnarray}
Expression~(\ref{8}) is the form from which one derives the usual
Feynman rules or
path integral which lead to anomaly of gauge symmetry and its
associated inconsistencies.

I have put a tilde on $U$ because I wish to claim that $\tilde U$
of~(\ref{8}) is not the solution for the Tomonaga-Schwinger
equation in the case of chiral gauge theory.  It would have been the
solution only if when considering the time derivative
\begin{eqnarray}
\partial_t\tilde U(t,t_0) &=&
\lim_{\epsilon\to 0}\frac{\tilde U(t+\epsilon,t_0)-
\tilde U(t,t_0)}{\epsilon}
\label{9}
\end{eqnarray}
one had ignored, as one often does, terms of {\it apparent}
order $O(\epsilon)$ or higher like
\begin{eqnarray}
\lim_{\epsilon\to 0} \frac{(-i)^2}{2!\epsilon}\int^{t+\epsilon}_t
\int^{t+\epsilon}_t{\cal T}[H_{int}(\tau)H_{int}(\tau')]d\tau d\tau'.
\label{10}
\end{eqnarray}
However, the time-ordered product is highly singular at equal time
arguments and may be able to reduce the term above to order
$O(\epsilon^0)$.  This is feasible, and is in fact the case, if the
singularity behaves as a delta function killing of one integration
of~(\ref{10}).
It is worth stressing that such singularity is more than the usual
statement of the indefiniteness of the time-ordered product at
equal times.

Wolfgang Pauli gave the example of a scalar field theory with
dervivative interactions wherein expression~(\ref{10}) does not
vanish~\cite{pauli}.  I now consider the example of
chiral gauge interactions as typified in the Chiral Schwinger Model.
\section*{Chiral Schwinger Model}
The Chiral Schwinger Model is a two dimensional abelian chiral
gauge theory.  It has the lagrangian density of~(\ref{1}), from which
follows the interaction Hamiltonian
\begin{eqnarray}
H_{int}(t) &=& g\int A_\mu J^\mu_L(x,t)dx,\nonumber\\
&=& g\int(A_0+A_1)J_{0L}(x,t)dx.
\label{8b}
\end{eqnarray}
We have used the metric $(1,-1)$ with $(t=x_0,x=x_1)$ and
$\gamma_5 = \gamma_0\gamma_1$ to get the last equation from
\begin{eqnarray}
J_{0L} &=& -J_{1L}.
\label{8aa}
\end{eqnarray}
For this model only expression of~(\ref{10}) is non-vanishing,
as we shall see.  All other terms of apparent order $O(\epsilon^2)$
or higher indeed vanish in the limit $\epsilon\to 0$.

The first term of the time-ordered product~(\ref{8a}), the
anti-commutator, does not
contribute to~(\ref{10}) but the other term involving commutator
does.  Of this commutator term, the contribution comes from the
commutator of fermion currents at different times~\cite{manton}
\begin{eqnarray}
[J_{0L}(x,\tau),J_{0L}(x',\tau')] &=& \frac{i\hbar}{2\pi}
\partial_x\delta(x-x'+\tau-\tau').
\label{11}
\end{eqnarray}
To evaluate~(\ref{10}) it suffices to consider
\begin{eqnarray}
K(t) &=& \lim_{\epsilon\to 0}\frac{i\hbar g^2}{2\pi}\frac{(-i)^2}
{2!\epsilon}\int dx
\int^{t+\epsilon}_t A_+(x,\tau) \partial_x Q(x,\tau,t) d \tau,
\label{12}
\end{eqnarray}
where $A_+ = A_0 + A_1$ and
\begin{eqnarray}
Q(x,\tau,t) &=& \frac{1}{2}\int dx'\int^{t+\epsilon}_t \sign(\tau-\tau')
\delta(x-x'+\tau'-\tau)A_+(x',\tau') d\tau'.
\label{12a}
\end{eqnarray}
{}From the identity
\begin{eqnarray}
\left(\partial_\tau - \partial_x\right)
\left(\frac{1}{2}\sign
(\tau-\tau')\delta(x-x'+\tau-\tau')\right) &=&
\delta(x-x')\delta(\tau'-\tau),
\label{14}
\end{eqnarray}
we have
\begin{eqnarray}
\left(\partial_\tau - \partial_x\right) Q(x,\tau,t) &=&
A_+(x,\tau),
\label{15}
\end{eqnarray}
as the delta functions have eliminated the integrations over $x'$
and $\tau'$.

Thus $K(t)$ survives the $\epsilon\to 0$ limit,
\begin{eqnarray}
K(t) &=& \frac{i\hbar g^2}{4\pi}\int A_+(x,t)\frac{\partial_1}
{\partial_0+\partial_1} A_+(x,t)dx.
\label{16}
\end{eqnarray}
(In our convention, $\partial_x = -\partial_1$.)
$K(t)$ spoils the Tomonaga-Schwinger equation for $\tilde U$
as claimed.

It is a simple combinatoric exercise to show that the true solution
of the Tomonaga-Schwinger equation is
\begin{eqnarray}
U(t,t_0) &=& {\cal T}\exp\left\{-i\int_{t_0}^t(H_{int}(\tau)
-iK(\tau))d\tau\right\}.
\label{17}
\end{eqnarray}
$K(t)$ leads to an extra term in the path-integral action
in comparison to the classical action (with appropriate
gauge-fixing term),
\begin{eqnarray}
{\cal S}_{extra} &=& -i\int K(t) dt,\nonumber\\
&=&  \frac{\hbar g^2}{4\pi}\int \left\{\left(\frac{\partial A}
{\Box}\right)\epsilon^{\mu\nu}\partial_\mu A_\nu + A_\mu\left(
g^{\mu\nu} - \frac{\partial^\mu\partial^\nu}{\Box}\right)A_\nu
\right.\nonumber\\
&&+ \left.A_1(A_0+A_1)\right\}dxdt.
\label{18}
\end{eqnarray}
Local counterterms are then needed to restore Lorentz invariance.

The Feynman rules for chiral gauge theory are clearly
not as naively expected.  From the gauge transformation of the
gauge fields, the first term on the right hand side of the
last equation has the right coefficient to
cancel out the anomaly~(\ref{4}).
The second term is the mass term for the tranverse gauge fields.
It is proportional to the coupling constant squared.

The gauge anomaly cancellation is exact to
all orders in perturbation theory thanks to the non-renormalisation
property of the coefficient of chiral anomaly~\cite{non-renormalisation}.
Gauge invariance is never lost.

To demonstrate that mass term for chiral fermions is admissible,
let me change the variables in the path integral as follows
\begin{eqnarray}
\psi' &=& \left[\exp\left(-ig\frac{\partial A}{\Box}\right)P_L+P_R\right]
\psi,\nonumber\\
\bar\psi' &=& \bar\psi\left[P_L +\exp\left(ig\frac{\partial A}{\Box}\right)
P_R\right].
\label{20}
\end{eqnarray}
Associated with this change of variables is the Fujikawa
jacobean~\cite{fujikawa} which cancels out the the first term of
the right hand side of~(\ref{18}), modulo local counterterms.  The
path-integral action can be written in the form of a space-time integral over
\begin{eqnarray}
-\frac{1}{4}F^2 +\frac{\hbar g^2}{4\pi}A_\mu^TA^{\mu T} +
\bar\psi'(i\dslash +g\Aslash^TP_L)\psi'+ m\bar\psi'\psi',
\label{21}
\end{eqnarray}
where $A_\mu^T = P_{\mu\nu}A^\nu$ is the tranverse part of $A_\mu$
projected out by the tranversality projector $P_{\mu\nu}$.
As the new fermionic fields~(\ref{20}) are {\it gauge invariant},
a mass term $m\bar\psi'\psi'$ can be added to the path-integral
action without spoiling its gauge invariance.  Such a mass term, not
being protected by any symmetry, will indeed be generated by radiative
corrections in general.
We thus need no scalar field to generate masses for
both gauge and fermionic fields with chiral interactions.

Also from~(\ref{21}) it can be seen that the gauge current that is
coupled to the gauge fields is
\begin{eqnarray}
C_\mu &=& P_{\mu\nu}J^\nu_L.
\label{22}
\end{eqnarray}
The tranversality
projector $P_{\mu\nu}$ renders the gauge current conserved automatically,
in compatibility with the equation of motion,
\begin{eqnarray}
\partial_\mu F^{\mu\nu} + \frac{\hbar g^2}{2\pi}A^{\nu T} &=&
gC^\nu.
\label{22aa}
\end{eqnarray}
One can then exploit the current conservation to prove the renormalisability
and unitarity for abelian gauge group even in dimensions higher than
two~\cite{tdk-mp}.

If one chooses the Lorentz gauge in~(\ref{21}),
\begin{eqnarray}
\partial A &=& 0,
\label{19}
\end{eqnarray}
then the usual, anomalous formulation of the Chiral Schwinger Model
is recovered but with mass terms for the fields.  From this we
can now see that the conventional
formulation although apparently anomalous is consistent because it is
the gauge-fixed form of a gauge-invariant theory and because
there is no ghost for abelian gauge group.  Such consistency has been
explicitly verified elsewhere~\cite{jackiw} (without fermion
mass term) and~\cite{ramallo} (with explicit fermion mass term).
\section*{Concluding remarks}
In four dimensions, one will have to consider, in addition
to~(\ref{10}), term of of order $O(\epsilon^2)$
\begin{eqnarray}
\frac{1}{3!\epsilon}\int^{t+\epsilon}_t\int^{t+\epsilon}_t
\int^{t+\epsilon}_t
{\cal T}[H_{int}(\tau)H_{int}(\tau')H_{int}(\tau'')]d\tau d\tau'd\tau''
\label{21a}
\end{eqnarray}
for abelian gauge group and even higher-order terms for
non-abelian groups.  With the former group, the gauge current can be
written in the form~(\ref{22}) which is obviously divergenceless,
facilitating the proof of renormalisability and unitarity~\cite{tdk-mp}.
With non-abelian groups, proof of such desirable properties is much
more involving but can be constructed.  I will present it somewhere
else.

Expressions~(\ref{10}) and~(\ref{21a}) are connected directly to the
one-loop anomalous Feynman graphs.  These expressions cancel
the current anomalies exactly to all orders thanks to the
non-renormalisation theorem of chiral anomalies.  Furthermore,
the non-vanishing of these terms highlights and clarifies the
relationship between chiral anomalies and the non-closure of
current commutators, and of double and triple commutators (corresponding
with time ordering of three and four currents respectively).  The
violation of the abelian-current Jacobi identity~\cite{levy}, being a
special case of the non-closure of double commutators at different times,
can now be seen as necessary to preserve chiral gauge invariance.
I suspect that the Malcev identity is likewise violated
for non-abelian chiral currents.  Consequently, we may appreciate
more the relevance of two and three cocycles~\cite{jackiw2}.

It should be emphasised that even though the above treatment
prevents chiral gauge invariance from going anomalous,
it has no consequence on the anomalies of global symmetries.
This is simply because the
Noether current of a global continuous symmetry does not
appear in the lagrangian density at all and hence cannot modify
the Feynman rules that render it anomalous in the first place.
Anomalies do exist and have important relevance to nature through
the breaking of global invariances.

The path-integral action of the Chiral Schwinger Model contains
non-local terms.  This is also a general feature for higher dimensions.
However, like the non-locality associated with the introduction of
Faddeev-Popov
ghosts, this is only in appearance.  The microcausality from vanishing
space-like commutators of physical fields are untouched.  In fact,
similar to the existence of ghost-free gauges in non-abelian theory,
one can fix a gauge such that all the non-local terms disappear.

On the other hand, with this new degree of non-locality
we might be able to avoid the theorem of~\cite{smith}.
The theorem is a kind of a no-go theorem stating that the only
(non-abelian) vector-meson theories with tree graphs with good
high-energy behaviour are Higgs theories.  Work in this direction
will be reported when it is completed.

Let me move to another no-go theorem, that of lattice chiral
fermions~\cite{nn}.  The quantisation of this paper proposes a way
around the theorem in that global axial can be
broken while chiral gauge symmetry is preserved~\cite{tdk-latt}.
This solution for lattice chiral fermions should also provide
an answer to the question raised in~\cite{banks}.  Numerical
simulations of a chiral gauge theory is in progress.

The existence of new Feynman rules can also be seen
through the Berry's phase of a path-integral quantisation of
chiral gauge theory~\cite{tdk-berry}.  Based on that work and the
work reported here, I also suspect that there is no anomaly associated
with homotopically non-trivial gauge transformations~\cite{witten}.

The answer to the title question is that there is nothing wrong with
``anomalous'' chiral gauge theory simply because it is not anomalous
at all.  Other questions can be raised here:
What are the implications on the electroweak
interactions?  Can the Higgsless mass generation be realised?

I wish to thank many peoples, Ian Aitchison, Richard Dalitz, Johnathan
Evans, Chris Griffin, Bruce McKellar and Brian Pendleton, for discussions
and support during the course of this work.  This work is partially
supported by the Australian Research Council.

\end{document}